\documentclass{aa} 

\usepackage{eso-pic}    
\usepackage{color}
	\definecolor{unipd}{HTML}{b5121b}
	\definecolor{cadmiumgreen}{RGB}{2,134,121}
	\definecolor{blue}{RGB}{0,32,169}
\usepackage{multirow}
\usepackage{booktabs}
\usepackage{relsize}
\usepackage{hyperref}
   \hypersetup{colorlinks=true, linkbordercolor=blue, breaklinks=true, citecolor=blue, filecolor=black, linkcolor=blue, urlcolor=cadmiumgreen}

\usepackage{graphicx}
\usepackage{txfonts}

\begin{document}

   \title{HR6819: a puffed-up stripped star system challenging stable mass transfer theory}

   \author{A. Picco
          \inst{1,3}
          \and
          P. Marchant\inst{2,1}
          \and
          H. Sana\inst{1,3}
          \and
          J. Bodensteiner\inst{4}
          \and
          T. Shenar\inst{5}
          \and
          A. J. Frost\inst{6}
          \and
          K. Deshmukh\inst{1}
          \and
          J. S. G. Mombarg\inst{7}
          \and
          D. Pauli\inst{1}
          \and
          R. Willcox\inst{1}
          \and
          A. Kemp\inst{1}
          }

   \institute{Institute of Astronomy, KU Leuven, Celestijnlaan 200D, 3001 Leuven,
        Belgium,\\
              \email{annachiara.picco@kuleuven.be}
         \and
         Sterrenkundig Observatorium, Universiteit Gent, Krijgslaan 281 S9, B-9000 Gent, Belgium,
         \and
         Leuven Gravity Institute, KU Leuven, Celestijnenlaan 200D, box 2415, 3001 Leuven, Belgium
         \and 
         Anton Pannekoek Institute for Astronomy, University of Amsterdam, Science Park 904, 1098 XH Amsterdam, The Netherlands
\and
The School of Physics and Astronomy, Tel Aviv University, Tel Aviv 6997801, Israel
         \and
         European Southern Observatory, Alonso de Cordova 3107, Vitacura, Casilla 19001, Santiago, Chile
         \and
         Universit\'e Paris-Saclay, Universit\'e de Paris, Sorbonne Paris Cit\'e, CEA, CNRS, AIM, 91191 Gif-sur-Yvette, France\
             }

   \date{\today}

 
  \abstract
  {HR6819 is the first system with a puffed-up low mass stripped star + a classical Be star whose nature has been confirmed by optical interferometry. The system exhibits the most extreme mass ratio (15.7 $\pm$ 1.1), the lowest stripped star mass (0.270 $\pm$ 0.056 $M_{\odot}$), and one of the shortest orbital periods (40.3266 $\pm$ 0.0016 days) compared to similar observed binaries. As a post mass transfer candidate, HR6819 offers a unique opportunity to test the physics of binary interaction, especially the fraction of mass accreted by the Be progenitor (the efficiency of the mass transfer) required to produce the system's extreme mass ratio.} 
   {This work aims to reconstruct the possible evolutionary history of HR6819 in the context of stable mass transfer via Roche Lobe overflow. We want to explore how the tight constraints on the system's total mass, mass ratio and orbital period are limiting the range of possible progenitors of the system. }
   {Based on analytical expectations for the orbital evolution, we build grids of \text{MESA} simulations designed to match the present-day orbital period and mass ratio of the system, with different mass transfer efficiencies from fully conservative to $50\%$ efficient.}
   {We show that evolution via stable mass transfer cannot explain the combined extreme current mass ratio and tight orbital period of the system. There is a limit on how extreme the post-mass transfer mass ratio of the progenitor binary can be at a fixed detachment period, and this limit is dependent on the efficiency of the mass transfer episode: the less efficient the mass transfer episode, the less extreme the mass ratio at detachment. Even in the case of fully conservative mass transfer, the most extreme mass ratio we can produce with binary evolution simulations is $q\sim11.5$ at $P\sim 40\:\mathrm{days}$, significantly below the observed value. We also show that the reported luminosities for each component significantly exceed the value expected from their mass. In particular, based on simple stripped star models, we find that the luminosity of the bloated stripped star requires a star of mass $\sim 0.7\:M_{\odot}$, over twice the measured value.}
  {Our work shows that the post-interaction properties of HR6819, especially its extreme mass ratio and orbital period, cannot be produced by stable mass transfer under standard assumptions.} 


   \maketitle
%

\section{Introduction}
HR6819 (HD 167128) is the first system that has been unambiguously confirmed by optical interferometry to contain a low mass stripped star caught in a puffed-up stage, and a rapidly rotating Be companion. Initially thought to be a hierarchical triple hosting an outer Be tertiary and an inner binary with a black hole + B-type star (\citealt{rivinius_naked-eye_2020}), it was later proposed (\cite{bodensteinerHR6819Triple2020}, \cite{el-badry_stripped-companion_2021}, \cite{Gies2020}) as one of the so-called black hole impostor systems (\citealt{bodensteinerDetectingStrippedStars2022}). Given the reflex motion of the Be star's emission lines with respect to the B-type star and the large mass ratio inferred from spectroscopy, they proposed that HR6819 was instead a post-mass-transfer binary with a low mass stripped star and a Be companion. This configuration has then been confirmed (\citealt{frostHR6819Binary2022}) by optical interferometry with long-baseline data taken by the Very Large Telescope Interferometer (VLTI, \citealt{Haubois2022}) with the GRAVITY instrument (\citealt{Gravity2017}) at high spectral resolution. This data clearly resolved a stripped object and the decretion disk surrounding the Be star, and gave a first estimate of a mass for the system ($3-7\:M_{\odot}$). More recently, further interferometric monitoring together with existing spectroscopy has allowed to constrain the full 3-dimensional orbit (\citealt{klementVLTIGRAVITYEnables2025}), revealing an extreme mass ratio of $m_{\mathrm{Be}}/m_{\mathrm{stripped}}= 15.7\pm 1.1$.

Stripped stars are stars that have lost most of their hydrogen-rich envelope to a companion and they are a common outcome of the evolution of close binary systems (\citealt{Sana_2012}). At low initial masses the removal of the envelope is mainly achieved through stable mass transfer (MT) via Roche Lobe overflow (RLOF) (\citealt{kippenhahn_entwicklung_1967}) or through common envelope evolution (\citealt{Paczynski1971}). Evolutionary models predict that, after the stripping, the star is a puffed-up naked helium core undergoing a contraction phase (e.g.,  \citealt{Claeys2011}). 
Subsequently, a long-lived phase of helium burning (10\% of the stellar lifetime) is expected, during which the stripped star will appear hot and bright in the UV (\citealt{gotberg_spectral_2018}) and challenging to observe with optical telescopes due to its much lower luminosity with respect to the companion. 
Due to the accretion of mass and angular momentum, the companion star is predicted to rapidly rotate, possibly producing a decretion disk aligned with the orbital plane (\citealt{Rivinius_2013}), therefore producing a Be companion.

The stripping phenomenon in binaries can occur at any mass, but their spectral appearance will strongly depend on the stripped star mass and its mass loss properties. It can produce Wolf Rayet stars (with masses $M \gtrsim 8\:M_{\odot}$, \citealt{crowther_physical_2007}, \citealt{Shenar2020}), intermediate-mass stripped stars ( $2\:M_{\odot}\lesssim M \lesssim 8\:M_{\odot}$, \citealt{droutDiscoveryMissingIntermediatemass2023}, \citealt{gotberg_stellar_2023}, \citealt{ludwigStrippedStarUltravioletMagellanic2025}) and OB-type subdwarfs (sdOBs, $M \lesssim 2\:M_{\odot}$, \citealt{heberHotSubdwarfStars2025}), depending on the metallicity of the environment. In particular, sdOBs are thought to be sustained by core or shell helium burning as they transition to the white dwarfs (WDs) sequence, though the lowest mass subdwarfs may evolve through this stage without any support by nuclear fusion (e.g. \citealt{Heber2003}). They are found in orbit with WDs (\citealt{kupfer_hot_2015}), Main Sequence (MS) stars (\citealt{van_roestel_discovery_2018}, \citealt{vos_composite_2018}, \citealt{schaffenrothHotSubdwarfsClose2022}) and Be stars. The sdOB companions to bright Be stars can be directly detected with UV spectroscopy (e.g. \citealt{wangDetectionCharacterizationBe+sdO2021}) and infrared interferometry (e.g. \citealt{klementCHARAArrayInterferometric2023}), a prototypical example being $\phi$ Per (\citealt{gies_hubble_1998}, \citealt{mourard_spectral_2015}). To date, targeted explorations confirmed roughly two dozens Be + sdOB binaries with orbital periods of the order of months (e.g. \citealt{klementCHARAArrayInterferometric2023}, \citealt{wang_orbital_2023}), which seems to support the stable MT formation scenario according to population synthesis studies (\citealt{han_origin_2003}). 

In HR6819, the stripped star is observed with a larger radius (hence, puffed-up) and a cooler effective temperature than normal helium-burning sdOBs, a stage that has been referred to as "bloated stripped star" and is possibly representative of the contraction stage prior to sdOBs. The bloated stage was traditionally expected to be short-lived (with a duration of the order of the thermal timescale of the stripped star, see e.g. \citealt{kippenhahn_entwicklung_1967}, \citealt{vandenheuvel1975}) and out of thermal equilibrium. However, evolutionary models have recently shown this stage to be near thermal equilibrium, and to last longer than expected (\citealt{el-badry_stripped-companion_2021}, \citealt{duttaEvolutionaryNaturePuffedup2024}), possibly solving the tension with the growing number of puffed-up stripped stars observed so far. In contrast to standard sdOBs, bloated stripped stars are detectable in the optical and their surface properties closely resemble those of B-type MS stars of much higher mass, making their identification challenging. A first example was the case of LB-1, thought to host a $\sim70\:M_\odot$ black hole (\citealt{liu_wide_2019}) and later proposed to hide a puffed-up stripped star with a mass of $\sim1.5\:M_\odot$ and a $\sim7\:M_\odot$ Be companion (\citealt{Irrgang2020}, \citealt{Shenar2020}). Joining HR6819 and LB-1, there is currently a handful of such low mass bloated stripped stars + Be star candidates (NGC 1850, \cite{Saracino2022}, \citealt{el-badry_ngc_2022}, \citealt{riviniusNewbornBeStar2025}), and a few more in the intermediate mass range (e.g. $\gamma$ Columbae, \citealt{Irrgang2022}, VFTS-291, \citealt{Villasenor2023}). 
Notably, the very low mass of the stripped star in HR6819 sits right at the lower boundary of possible sdOB masses (\citealt{Zhang2009}, \citealt{Bauer2021}, \citealt{Arancibia2024}).

The rapidly rotating Be companion in HR6819 is thought to be a representative of a post-MT spun up accretor, supporting the idea that Be stars are formed more commonly in binaries (e.g. \citealt{Demink2013}, \citealt{Shao2014}) than in single-star scenarios (\citealt{Ekstrom2008}, \citealt{Granada2013}, \citealt{Hastings2020}). However, the amount of mass that such an accretor can retain during the MT (the efficiency) is still unconstrained. The traditional view is that, after just a few percent mass growth, the mass gainer would spin up to critical rotation (\citealt{Packet1981}), therefore stopping the accretion. Additionally, rotationally enhanced winds can make the MT very inefficient, unless strong tides are preventing the spin up to critical (\citealt{Langer2003}). 
State of the art evolutionary models usually account for tides and rotationally limited accretion, but the low MT efficiencies are challenged by the observations of sdOB + Be binaries, for which there is evidence for substantial accretion (\citealt{Pols2007}, \citealt{Bao2025}, \citealt{Lechien2025}, \citealt{Xu2025}). An alternative explanation may be that the specific angular momentum of the accreted material is only a fraction of the respective Keplerian momentum (\citealt{Paczynski1971}), but this effect has so far not been explored.

Systems like HR6819 are of special interest, as they can be representative of the immediate post-MT stage, providing a window into a previously unprobed stage of binary evolution and the opportunity to test MT theory. Compared to the other well characterised puffed-up stripped star + Be star systems, HR6819 exhibits the most extreme mass ratio ($m_{\mathrm{Be}}/m_{\mathrm{stripped}}= 15.7\pm 1.1$), the lowest stripped star mass ($m_{\mathrm{stripped}}=0.270\pm 0.056\:M_{\odot}$) and one of the shortest orbital periods ($40.3266 \pm 0.0016$ days), being also one of the best constrained systems in terms of errors. 
In this paper, we use the refined constraints on the dynamical masses from combined interferometry and spectroscopy (\citealt{klementVLTIGRAVITYEnables2025}) to model the evolution of a progenitor of HR6819, Sec. \ref{sec:methodology}. In Sec. \ref{sec:results}, we show that evolution via stable MT cannot explain the combined extreme current mass ratio and tight orbital period of the system. We further discuss the generality of the problem and provide our conclusions in Sec. \ref{sec:discussion}. 

\section{Methodology}\label{sec:methodology}

We use version 24.03.1 of Modules for Experiments in Stellar Astrophysics
\citep[\text{MESA,}][]{Paxton2011, Paxton2013, Paxton2015, Paxton2018, Paxton2019, Jermyn2023} to model the evolution of a progenitor binary to HR6819. A detailed description of our simulation setup is provided in Appendix \ref{sec:appMESA} and the input files are uploaded in Zenodo\footnote{\href{https://doi.org/10.5281/zenodo.15662480}{doi.org/10.5281/zenodo.15662480}} for reproducibility.

We start our simulations from two ZAMS stars in orbit with an initial period $P_{\mathrm{i}}$ and with initial masses $m_{\mathrm{stripped,i}}$ and $m_{\mathrm{Be,i}}$ for HR6819's stripped and Be star progenitors, respectively. We investigate the stable MT formation scenario, in which the initially more massive (primary) star is stripped of its envelope by a MT episode onto the initially less massive companion (secondary). We define the binary mass ratio $q$ as 
\begin{equation}
    q\equiv\dfrac{m_\mathrm{Be}}{m_\mathrm{stripped}}\:.
\end{equation}  
We compute the MT rates with the contact scheme from \cite{marchant_new_2016}. During the MT, the secondary accretes mass and becomes more massive than the primary, thus proceeding faster in its evolution and expanding within its own Roche Lobe. When MT from the secondary onto the primary (inverse MT) is initiated, we stop the simulation. We also single out the simulation where the accreting star expands beyond the volume-equivalent radius at the L2 Lagrangian point (L2 limit) following Eq. 2 in \cite{marchant_new_2016}, and exclude them from the analysis as they are assumed to become unstable.

To determine the initial conditions for our simulations, $P_{\mathrm{i}}$, $m_{\mathrm{stripped,i}}$ and $m_{\mathrm{Be,i}}$, we consider mass and angular momentum conservation during stable MT (\citealt{Soberman1997}, but see also \citealt{tauris2023physics}). We define the efficiency, $\epsilon$, of the MT as 
\begin{equation}\label{eq:epsilon_uguale_uno_meno_beta}
\epsilon=\dfrac{\Delta m_{\mathrm{accretor}}}{\Delta m_{\mathrm{donor}}}\equiv 1-\beta\:,
\end{equation}
where $\Delta m_{\mathrm{accretor}}$ is the fraction of accreted mass when the donor transfers $\Delta m_{\mathrm{donor}}$. Here we allow for a fraction $\beta$ of the transferred mass to leave the system with the specific angular momentum of the accreting star (commonly referred to as isotropic re-emission). Note that $\beta=0$ is the limit of conservative MT and $\beta=1$ describes fully inefficient accretion.
Mass conservation given a certain efficiency $\epsilon$ imposes the following relation between the initial total mass, $m_{\mathrm{tot,i}}=m_{\mathrm{Be,i}}+m_{\mathrm{stripped,i}}$, and the total mass at a reference point, $m_{\mathrm{tot,r}}$:
\begin{equation}\label{eq:total_mass}
    \dfrac{m_{\mathrm{tot,i}}}{m_{\mathrm{tot,r}}}=\left(\dfrac{1+q_{\mathrm{i}}}{1+q_{\mathrm{r}}}\right)\left(\dfrac{1+\epsilon q_{\mathrm{i}}}{1+\epsilon q_{\mathrm{r}}}\right)^{-1}\:.
\end{equation}
Finally, we assume circular orbits and constant fractions $\epsilon$ and $\beta$ along the MT, and we ignore other mechanisms for angular momentum loss from the system. Our models do not include rotation, since including a rotationally-limited accretion would reduce the MT efficiency and possibly mimic a very inefficient MT episode, in contrast with the observed sample of sdOBs + Be systems. For this reason, we rather want to explore different fixed efficiencies, to make a statement about how conservative the MT needs to be to reproduce the orbital properties of the system. Under these assumptions, one can derive the following map between the initial orbital period, $P_\mathrm{i}$, and the period at any reference point, $P_{\mathrm{r}}$, along the MT:
\begin{equation}\label{eq:shrinkage}
  \dfrac{P_{\mathrm{i}}}{P_{\mathrm{r}}}=\left(\dfrac{q_{\mathrm{r}}+1}{q_{\mathrm{i}}+1}\right)^{-\frac{3\beta\epsilon}{\epsilon(\epsilon-1)}-1} \left(\dfrac{\epsilon q_{\mathrm{r}}+1}{\epsilon q_{\mathrm{i}}+1}\right)^{\frac{3\beta}{\epsilon(\epsilon-1)}-5}\left(\dfrac{q_{\mathrm{r}}}{q_{\mathrm{i}}}\right)^3\:.
\end{equation}

   \begin{figure}
   \centering
   \includegraphics[width=0.5\textwidth]{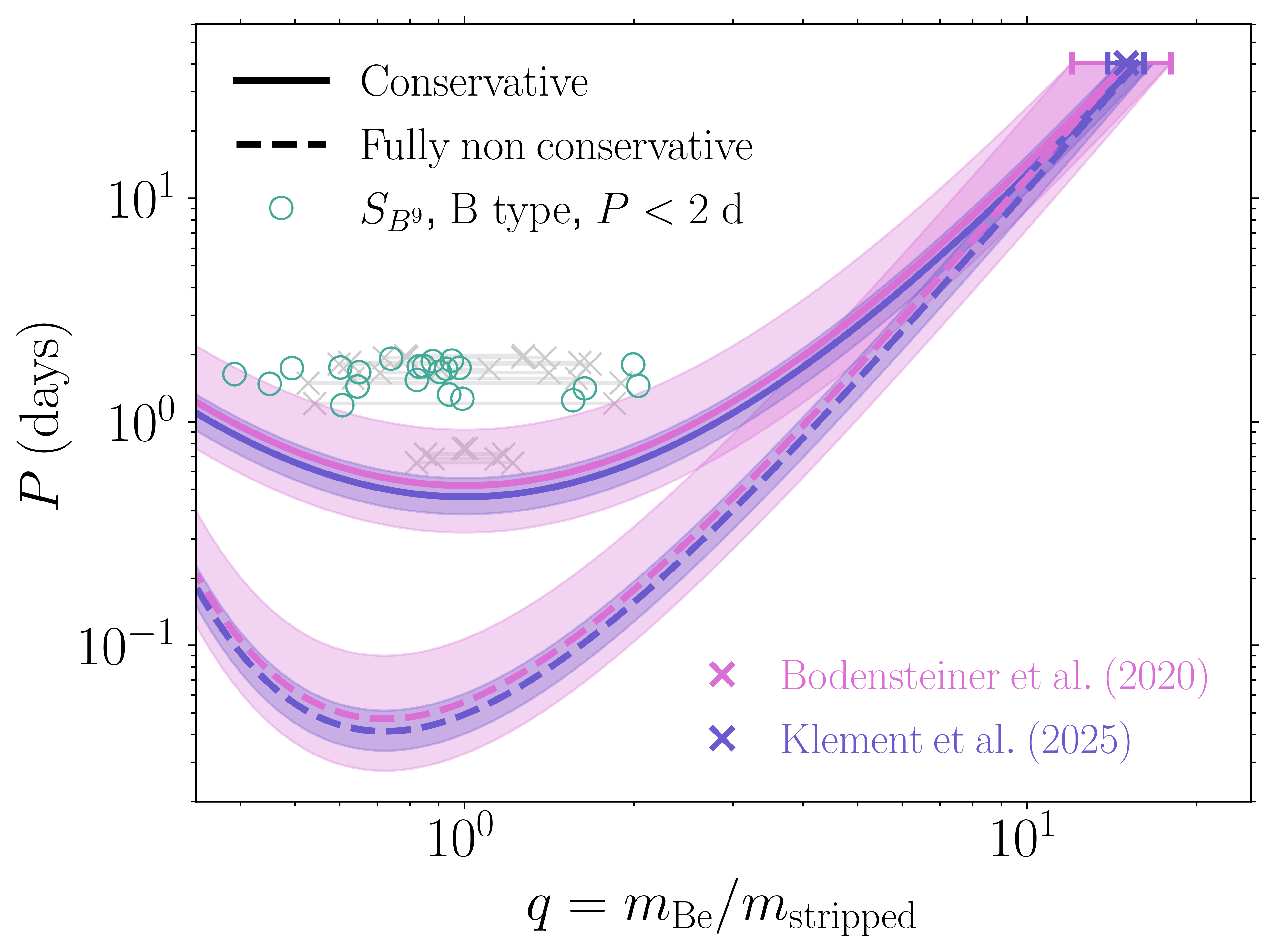}
      \caption{Orbital period $P$ of the binary system as a function of mass ratio $q$, assuming a present-day period of $P_{\mathrm{r}}\sim 40$ days and the two estimates of the current mass ratio from spectroscopy alone (purple, \citealt{bodensteinerHR6819Triple2020}, $15\pm 3$) and combined spectroscopy and optical interferometry (pink, \citealt{klementVLTIGRAVITYEnables2025}, $15.7\pm 1.1$), with their respective uncertainties (lighter colored bands around the respective line). HR6819 properties from these two works are also reported as scatter points, with their errorbars on the mass ratio (the errorbars on the period are too small to be visible). MT is assumed to be either fully conservative (solid) with $\epsilon=1$ or fully non conservative (dashed) with $\beta=1$. Circles are a selection of detached and semidetached systems from the S$_{\mathrm{B}^9}$ catalogue, same as Figure 15 from \cite{bodensteinerHR6819Triple2020}, see also their Appendix D.
              }
         \label{fig:theory}
   \end{figure}

For the references $q_{\mathrm{r}}$, $P_\mathrm{r}$ and $m_{\mathrm{tot,r}}$, we consider today's values $q_{\mathrm{today}}$, $P_\mathrm{today}$ and $m_{\mathrm{tot,today}}$ constrained by combined spectroscopy and optical interferometry (\citealt{klementVLTIGRAVITYEnables2025}), summarized in Table \ref{tab:ref_values}. We show in Fig. \ref{fig:theory} the result of evaluating Eq. \ref{eq:shrinkage} for $q_{\mathrm{r}}=q_{\mathrm{today}}$, compared with the mass ratio as given by spectroscopy alone (\citealt{bodensteinerHR6819Triple2020}). The binary system starts the MT with a mass ratio $q<1$ and, as the MT proceeds, the Be star becomes the more massive object and $q>1$, i.e., the diagram follows the time from left to right. The current constraints on the mass ratio of HR6819 from the updated orbital solution limit the possible initial orbital periods to $P_{\mathrm{i}}\lesssim$ 1.25 day, when the MT is assumed to be conservative. Fully non conservative MT requires even tighter initial configurations $P_{\mathrm{i}}\lesssim 4\:\mathrm{h}$. Considering also the double-line spectroscopic binaries with a period of less than 2 d in the $S_{B^9}$ catalogue (\citealt{pourbaix_sb9_2004}), we see that none of these systems could serve as the progenitor of HR6819, given the tight constraints on the present-day mass ratio from \cite{klementVLTIGRAVITYEnables2025}. The possible initial configurations can be wider if one allows for a less extreme value of $q_{\mathrm{r}}$ to be reached at the present-day period $P_{\mathrm{today}}$. In particular, \cite{bodensteinerHR6819Triple2020} modeled a progenitor system with masses $6\:M_{\odot}$ and $2\:M_{\odot}$ and a period of $2\:\mathrm{days}$. Their model undergoes conservative case A MT and reaches a mass ratio at detachment as high as $m_{\mathrm{Be}}/m_{\mathrm{stripped}}\sim12.5$, which falls within the uncertainty range of the mass ratio estimate from spectroscopy alone, but is inconsistent with the tighter constraints from \cite{klementVLTIGRAVITYEnables2025}.

\begin{table}
    \centering
    \renewcommand{\arraystretch}{1.2} 
    \setlength{\tabcolsep}{10pt} 
    \caption{Properties of HR6819}
    \begin{tabular}{cc}
        \toprule
        $q_\mathrm{today}$ & $15.7 \pm 1.1$ \\
        $P_{\mathrm{today}}\:(\mathrm{days})$ & $40.3266 \pm 0.0016$ \\
        $m_{\mathrm{tot,today}}\:(M_{\odot})$ & $4.28 \pm 0.36$ \\
        $m_{\mathrm{stripped,today}}\:(M_{\odot})$ & $0.270\pm0.056$ \\
        $m_{\mathrm{Be,today}}\:(M_{\odot})$ & $4.03\pm 0.34$ \\
        \bottomrule
    \end{tabular}
    \vspace{0.6em}
    \tablefoot{Reference values for HR6819 used in this work to set up the \text{MESA} simulations. This table is a subset of table 3 of \cite{klementVLTIGRAVITYEnables2025}, where we use their circular solution derived from astrometry + radial velocity fitting. The individual masses $m_{\mathrm{stripped,today}}$ and $m_{\mathrm{Be,today}}$ are not directly used in the setup, but reported for completeness. Note: their table uses the inverse definition of the mass ratio $q = m_{\mathrm{stripped}}/m_{\mathrm{Be}}=0.0638\pm 0.0045$.}
    
    \label{tab:ref_values}
\end{table}

At fixed efficiencies $\epsilon$ and $\beta$, Eq. \ref{eq:shrinkage} and Eq. \ref{eq:total_mass} combined give us the possible initial values for our simulations. However, fixing $q_{\mathrm{r}}=q_{\mathrm{today}}$ and $P_{\mathrm{r}}=P_{\mathrm{today}}$ does not guarantee that the binary simulation will reach the target mass ratio, as the system can detach before or after reaching $q_{\mathrm{today}}$, or MT might become unstable. As we will show in Sec. \ref{sec:results}, if we fix both $P_{\mathrm{r}}=P_{\mathrm{today}}$ and $q_{\mathrm{r}}=q_{\mathrm{today}}$, we find no solution due to these reasons.

\begin{figure}
   \centering
   \includegraphics[width=0.5\textwidth]{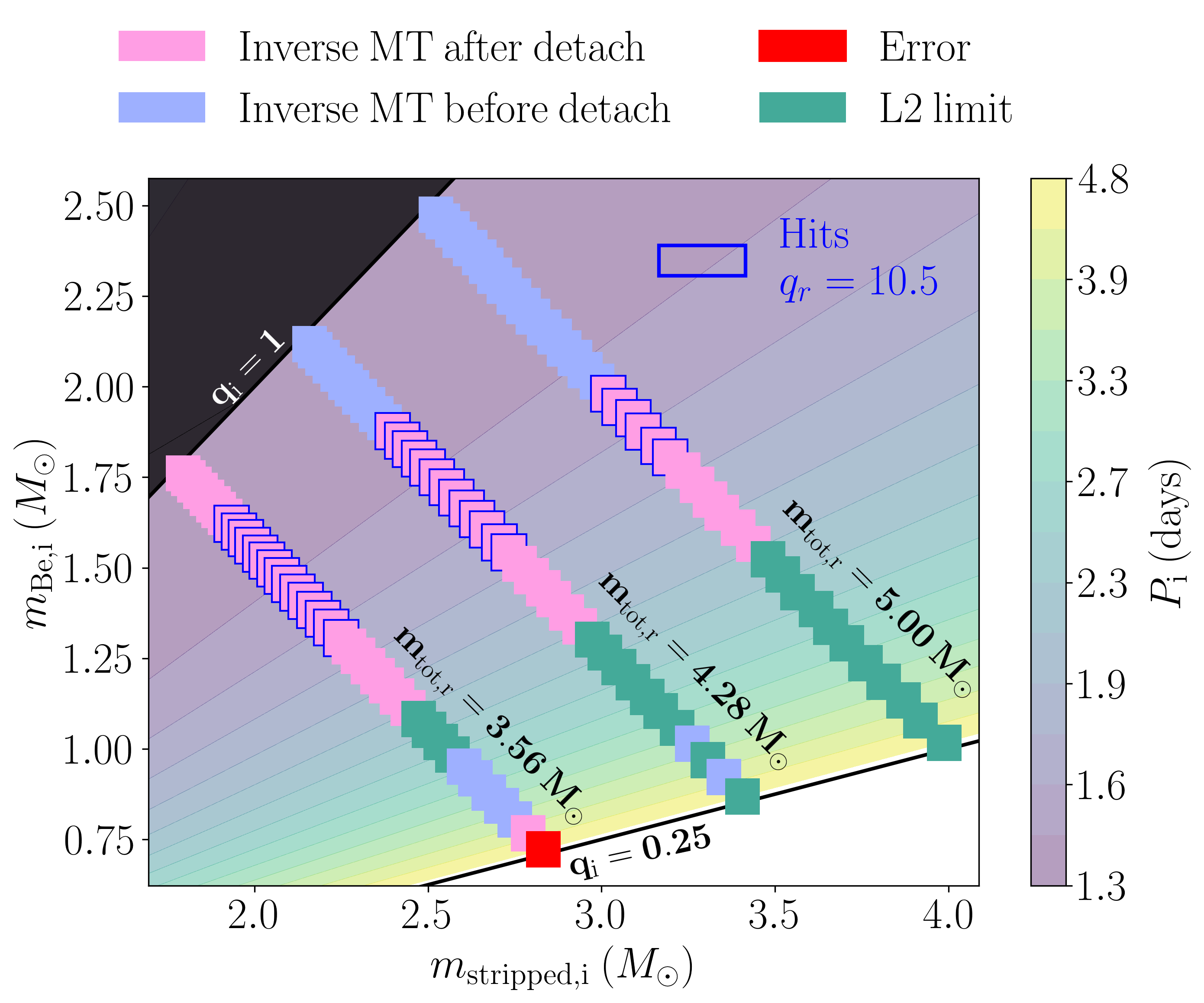}
      \caption{Contour levels of initial orbital periods $P_{\mathrm{i}}$ allowed by stable MT with $\epsilon=1$ efficiency, if the period at detachment is imposed to be $P_{\mathrm{r}}=P_{\mathrm{today}}$ from Table \ref{tab:ref_values} and the reference mass ratio is $q_{\mathrm{r}}=10.5$. We populate the lines of constant total mass at present-day with \text{MESA} simulations, represented by their outcome with color-coded pixels. If these simulations reach a mass ratio at detachment $q_{\mathrm{detach}}$ such that $q_{\mathrm{detach}}\geq q_{\mathrm{r}}$, we mark them as hits (blue edges). This is shown as a function of stripped and Be star initial masses, $m_{\mathrm{stripped,i}}$ and $m_{\mathrm{Be,i}}$. The black region indicates initial mass ratios $q_{\mathrm{i}}>1$, and the solid black line delimiting the base of the contour plot corresponds to $q_{\mathrm{i}}=0.25$, i.e., the lowest mass ratio we evolve.
              }
         \label{fig:grid}
   \end{figure}

Given these considerations, we produce simulations of progenitor systems exploring different MT efficiencies $\epsilon$ and different target mass ratios $q_{\mathrm{r}}$, always keeping the reference period at $P_{\mathrm{r}}=P_{\mathrm{today}}$.  Figure \ref{fig:grid} showcases our determination of $P_{\mathrm{i}}$, $m_{\mathrm{stripped,i}}$ and $m_{\mathrm{Be,i}}$ for one of our grids. In this example, we show the results of fixing $q_{\mathrm{r}}$ to 10.5, as an illustration. We compute $P_{\mathrm{i}}$ from Eq. \ref{eq:shrinkage} at 45 values of initial mass ratios $0.25 \leq q_{\mathrm{i}} \leq 0.95$ such that Eq. \ref{eq:total_mass} is also fulfilled if the model reaches $q_{\mathrm{today}}$. We do this for the median value of today's total mass ($4.28\: M_{\odot}$) and repeat it for the $\pm 2\sigma$ values, so that the pixels in Fig. \ref{fig:grid} are lining along the three diagonals of constant $m_\mathrm{tot,r}$.

\section{Results}\label{sec:results}
The example grid in Fig. \ref{fig:grid} shows several possible outcomes. We find that the smaller the initial mass ratio, the easier it is for the primary to overflow beyond L2 and break the stable MT scenario; furthermore, the tighter the initial orbital period and the closer the mass ratio to unity, the easier it gets for the binary to initiate inverse MT from the secondary onto the primary. If inverse MT happens before detachment, the secondary overfills its own Roche Lobe before the primary could be stripped of its envelope. Therefore, the system is expected to undergo a contact phase and we do not consider it in the analysis, as such systems typically evolve into a merger (see e.g., \citealt{Fabry2025}). In case inverse MT happens after successful detachment, the primary is a stripped star when the secondary overfills its own Roche Lobe. By construction, these systems will evolve during the MT along a similar curve as in Fig. \ref{fig:theory}. If the mass ratio at detachment, $q_{\mathrm{detach}}$, is such that $q_{\mathrm{detach}}\geq q_{\mathrm{r}}$, the stable MT episode has succeeded in producing a stripped star in a system with a mass ratio as high as $q_{\mathrm{r}}$ at $P_{\mathrm{today}}$. Depending on their mass ratio at the onset of the inverse MT, such systems may undergo a merger leading to a more massive stripped star, similarly to the scenario proposed for the quasi-Wolf Rayet star HD 45166 (\citealt{Shenar2023}).

We first produced a grid similar to that in Fig. \ref{fig:grid}, with initial conditions determined by $q_{\mathrm{r}}=q_{\mathrm{today}}$ from Table \ref{tab:ref_values} and MT efficiencies $0.50 \leq \epsilon \leq 1.00$, spaced by $\Delta\epsilon=0.05$ (we do not consider $\epsilon$ values lower than 0.5 for computational reasons). We found no solution for the median of the present-day mass ratio $q_{\mathrm{r}}=15.7$, as anticipated from Fig. \ref{fig:theory}: to produce HR6819's extreme mass ratio with stable MT, very tight initial orbital configurations are required, and all simulations experienced either L2 overflow or inverse MT before forming a stripped star. Similarly, no solutions were found for the $-2\sigma$ or $-3\sigma$ values of $q_{\mathrm{r}}=13.5$ or $12.4$, respectively. 

   \begin{figure}
   \centering
   \includegraphics[width=0.5\textwidth]{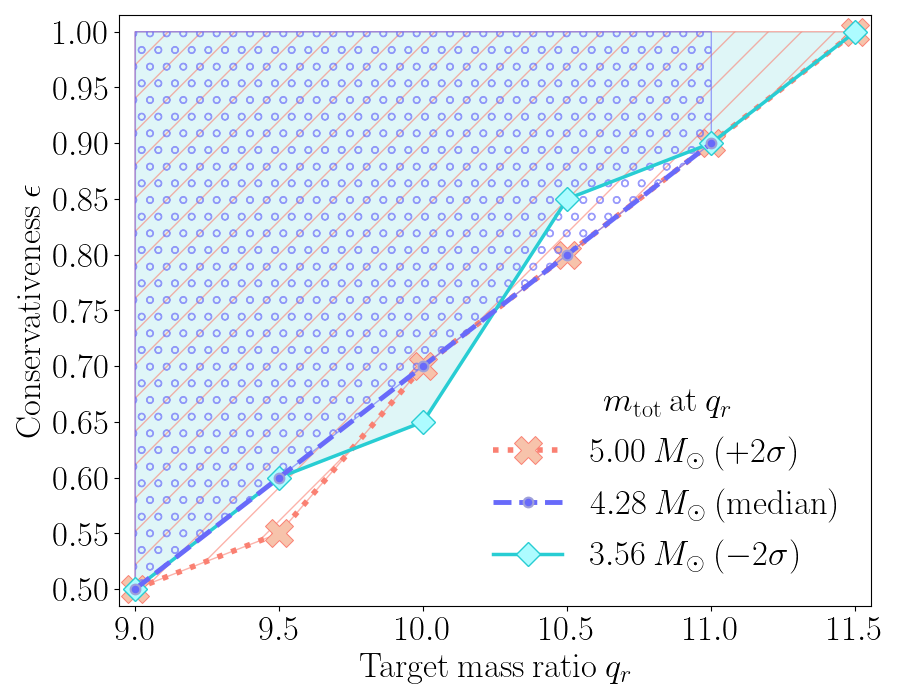}
      \caption{Summary plot of our grid exploration for different conservativeness degree $\epsilon$ and different target mass ratios $q_{\mathrm{r}}$. The area is filled where at least a solution with $q_{\mathrm{detach}}\geq q_{\mathrm{r}}$ is found, for three different values of the present-day total mass $m_{\mathrm{tot,r}}$: median (violet, with circular hatching); $-2\sigma$ value (cyan, with shading); $+2\sigma$ (salmon, with diagonal hatching). The scatter points along the respective lines indicate where the minimum value of $\epsilon$ that gives a solution is found.
              }
         \label{fig:Epsilon}
   \end{figure}

Since we found no progenitor system that can reach HR6819's extreme mass ratio at the present-day period, we investigated how extreme the mass ratio $q_{\mathrm{detach}}$ can get via stable MT, and the dependence of this limit on the conservativeness of the MT episode. For this purpose, we explored lower values for the target mass ratio $q_{\mathrm{r}}$. We produced several grids similar to Fig. \ref{fig:grid}, with $9.0\leq q_{\mathrm{r}}\leq 12.0$, spaced by $\Delta q_{\mathrm{r}}=0.5$. The results of the exploration are summarized in Fig. \ref{fig:Epsilon}. We show that to form a system with a more extreme mass ratio at period $P_{\mathrm{today}}$, one needs a more conservative stable MT episode. 
In particular, we find that the most extreme mass ratio that can be produced consistently with the present-day period $P_{\mathrm{today}}$ is $q_{\mathrm{r}}\sim 11.5$ (with a small dependence on the total mass of the system). This mass ratio is well below the $3\sigma$ lower limit for HR6819 given in Table \ref{tab:ref_values}. 

\begin{figure*}
   \centering
   \includegraphics[width=1.0\textwidth]{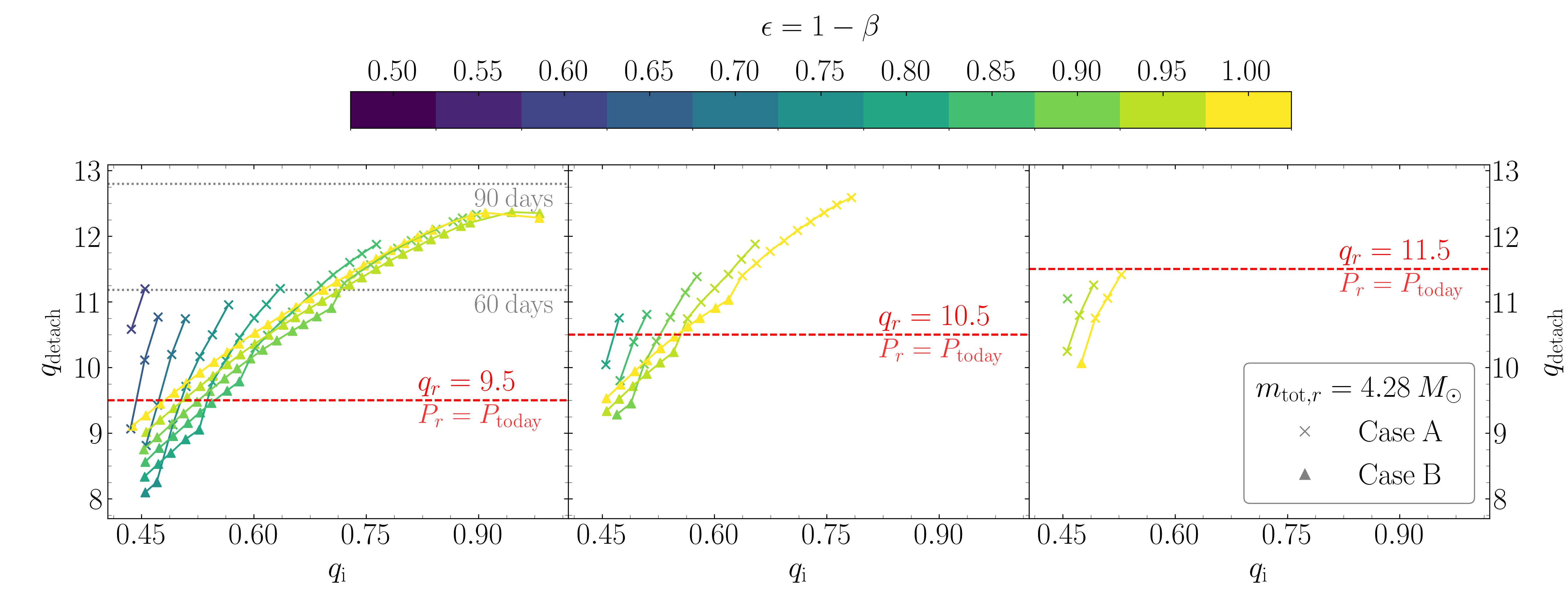}
   \caption{Detachment mass ratio as a function of initial mass ratio for a selection of our results. The three panels refer to three different grids for target mass ratio $q_{\mathrm{r}}$ as indicated by the dashed red line, which also corresponds to a detachment period of $P_{\mathrm{r}}=P_{\mathrm{today}}$, by construction. Different MT efficiencies are color-coded according to the colorbar, and the scatter points mark case A (x marker) and case B (triangle marker) MT. The systems have total mass $m_{\mathrm{tot,r}}=4.28\:M_{\odot}$ (median for HR6819) at $q_{\mathrm{r}}$. In the first panel, we also show in gray the detachment ratios corresponding to periods of $60$ and $90\:\mathrm{days}$ as a reference.}
              \label{fig:resolution}%
    \end{figure*}

Figure \ref{fig:resolution} presents a selection of our results for the detachment mass ratio at $m_{\mathrm{tot,r}}=4.28\:M_{\odot}$. For each initial mass ratio, we show the value of $q_{\mathrm{detach}}$ for systems that undergo inverse MT after successful detachment and stripping of the primary. The three panels show that increasing the target $q_{\mathrm{r}}$ makes it harder for the stable MT episode to fully strip the primary before inverse MT, due to tighter initial orbital configurations. Given the dependence of the initial period on the target mass ratio, the type of MT we expect (case AB or case B) is also influenced.
Lastly, the systems that undergo detachment with $q_{\mathrm{detach}}>q_{\mathrm{r}}$, have, by construction, a longer detachment orbital period than $P_{\mathrm{today}}$, as dictated by Eq. \ref{eq:shrinkage}. These systems show that stable MT can produce detachment mass ratios as high as $m_{\mathrm{Be}}/m_{\mathrm{stripped}}\sim 13$. These systems would have a period of $P\sim 90\:\mathrm{days}$, comparable with the observed sample of Be + sdOB binaries, but not compatible with HR6819's present-day period.

   \begin{figure}
   \centering
   \includegraphics[width=0.5\textwidth]{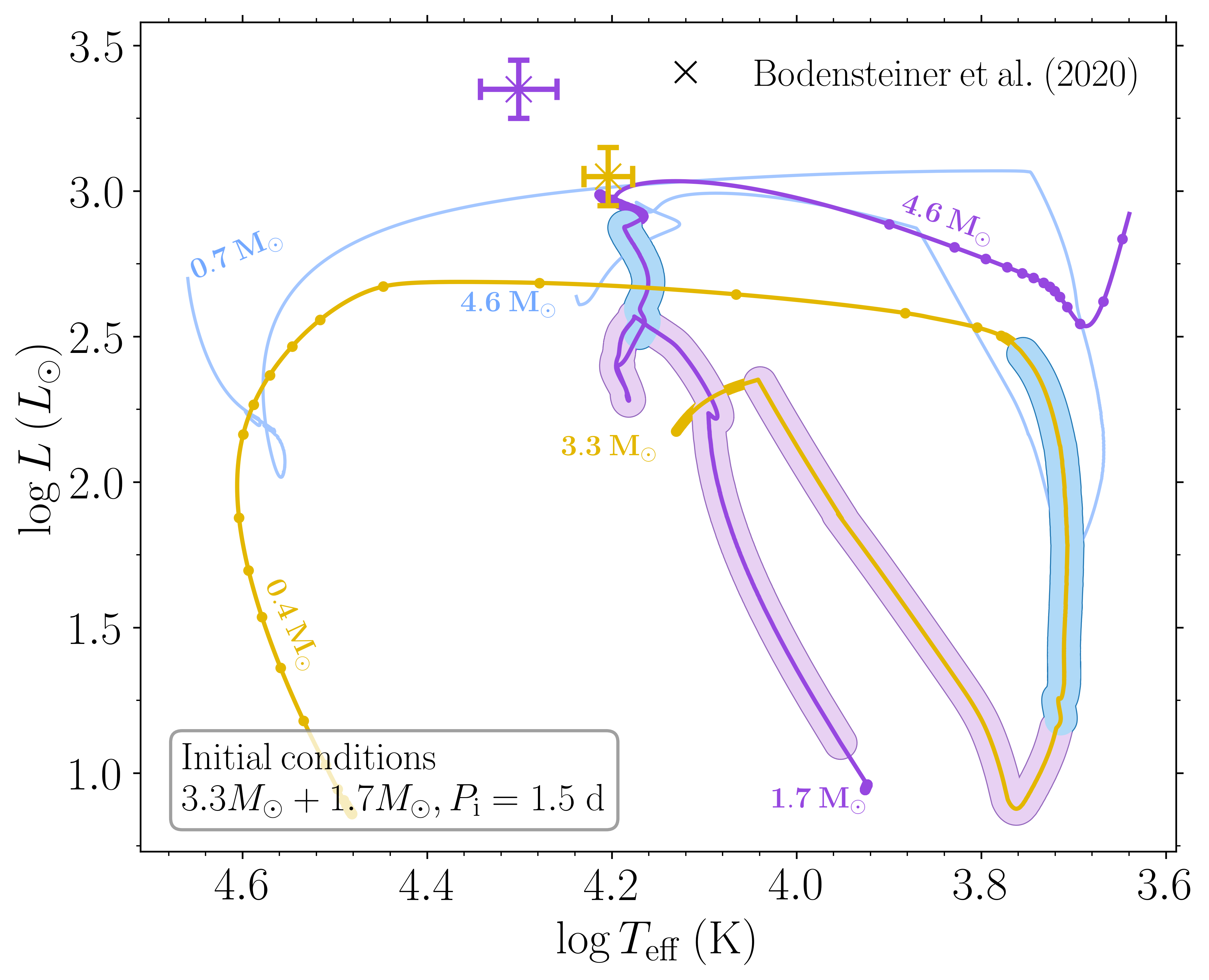}
      \caption{HR diagram showing the evolution of a system composed of a primary (the stripped star progenitor) with $m_{\mathrm{stripped,i}}=3.3\:M_{\odot}$ and a secondary (the Be star progenitor) with $m_{\mathrm{Be,i}}=1.7\:M_{\odot}$, in orbit with initial period $P_{\mathrm{i}}=1.5\:\mathrm{days}$. The system evolves via fully conservative MT. Dots along the tracks are separated by 0.1 Myr of evolution. Purple and blue highlighted regions correspond to mass transfer on the MS (case A) and after the MS (case AB), respectively. The scatter points mark the observed locations of HR6819 according to \cite{bodensteinerHR6819Triple2020}, with $1\sigma$ error bars. An evolutionary track for a donor star (ZAMS mass $4.6\:M_{\odot}$) undergoing case B mass transfer is shown in blue as it crosses HR6819's observable at its median. The ZAMS and stripped stars' masses for every track are also reported along the track with corresponding colors.}
         \label{fig:HR}
   \end{figure}

\section{Discussion and conclusions}\label{sec:discussion}
We calculated the evolution via stable MT of progenitors of stripped star systems and showed that it is not possible to produce mass ratios as extreme as the one measured for HR6819 from combined spectroscopy and interferometry. The initial orbital configurations of possible progenitors of HR6819 are too tight and would engage in a) MT from the secondary onto the primary before the complete stripping of the primary, b) L2 Lagrangian point overflow, depending on their initial mass ratio. High MT efficiency allows for larger initial periods and can help to overcome such difficulties; however, there is a limit to how extreme the mass ratio can get at a fixed post-interaction orbital period. A system that has evolved through stable MT can host a stripped star and its companion in an orbit of $P\sim40$ days with a mass ratio at most as high as $m_{\mathrm{Be}}/m_{\mathrm{stripped}}\sim11.5$. Conversely, the channel can produce mass ratios as high as $\sim 13$, i.e., possibly right at the $3\sigma$ limit of the observed value, but such a system would have a much wider orbital period ($P\sim 90$ days) than that of HR6819. 

Figure \ref{fig:HR} shows an example of a binary that, after the stripping of the primary via conservative MT, detaches with mass ratio $q\simeq 11.65$ at $P\simeq 41.79\:\mathrm{days}$. By construction, this system has a total mass at detachment of $\simeq 5\:M_{\odot}$ (the $+2\sigma$ value for HR6819's total mass), and crosses the target mass ratio $q_{\mathrm{r}}=11.5$ at $P_{\mathrm{r}}=P_{\mathrm{today}}$. The stripped star mass at detachment is $\simeq 0.4\:M_{\odot}$, and the secondary has become a $\simeq 4.6\:M_{\odot}$ star still on its MS. These masses are much too low to cross HR6819's observed values on the HRD, despite our specific choice to match the higher $2\sigma$ limit for the total mass of the system. To alleviate this problem, one could consider non conservative MT, that would allow for even higher initial masses of the components; however, our work shows that non conservative MT requires tighter initial orbital configurations and is not able to reproduce the mass ratio at the observed period. 

Furthermore, the present-day luminosity and effective temperature of HR6819 cannot be matched with a post-stable MT system, as both binary components would need to have a much higher ZAMS mass given the current total mass constraint. To visualize this, we produced evolutionary tracks for donor stars of different masses undergoing case B fully conservative MT and found an appropriate ZAMS mass that would explain HR6819's stripped star, see Fig. \ref{fig:HR}. We found that our model with ZAMS mass of $\sim 4.6\:M_{\odot}$ would have the luminosity and effective temperature of HR6819 on the contraction stage, but with a stripped star mass as high as $\sim 0.7\:M_{\odot}$. Such a stripped star mass is too high given the observed $m_{\mathrm{stripped}}=0.270\pm 0.056\: M_{\odot}$ (see Table \ref{tab:ref_values}). We use the BONNSAI\footnote{The BONNSAI web-service is available at \href{https://www.astro.uni-bonn.de/stars/bonnsai/}{www.astro.uni-bonn.de/stars/bonnsai}.} tool (\citealt{Schneider_2014}) to infer the appropriate mass of the secondary given its HRD properties, after its accretion from the primary: this would need to be a $ 7^{+ 0.52}_{-0.62}\:M_{\odot}$ star on its MS. This is expected as the mass gainer is enriched with helium and has a higher luminosity than a single star model of similar mass.

As our results from conservative stable MT are inconsistent with HR6819's observational constraints, one should think of possible ways to alleviate the tension. We expect convective overshooting to have an impact: a smaller core would lead to a more extreme mass ratio at detachment, as the donor star will then form a smaller helium core and thus, a smaller stripped star mass. Since we used a step-overshooting scheme of the hydrogen burning convective core, extending its size by $0.1$ pressure scale heights (see also Appendix \ref{sec:appMESA}), we explored also a much lowered value of $0.01$ as a parameter variation, and present the results in Appendix \ref{app:variation}. In the case of lowered overshooting parameter, we found that systems with mass ratios at most as high as $q\sim 12$ at $P_{\mathrm{today}}$ can be formed via stable MT. We therefore conclude that the tension cannot be solved with a lower overshooting parameter. 

Our assumption of fixed accretion efficiency throughout the MT episode might also have a significant impact on our outcomes. In HR6819, the Be star component is rotating near-critically, and analytical considerations predict that during the MT episode such a critical rotation rate is reached soon after the onset of RLOF, after accreting a few percent of its mass with the specific angular momentum of its surface (\citealt{Packet1981}). In this case, part of the momentum exchanged during RLOF feeds a non-zero spin angular momentum of the accretor instead of being transferred to the orbit. At this point, the spun-up star could keep accreting mass while maintaining its rotation thanks to a braking mechanism (\citealt{Deschamps2013} and references therein). Alternatively, mass may be lost from the system by feeding circumbinary outflows (e.g. \citealt{Lu2023}) or via localized radiation-driven winds from spots of direct-impact accretion (\citealt{vanRensbergen2008}). The occurrence of either of the two cases, and the subsequent degree of non conservativeness they might introduce, are major uncertainties that we are not accounting for in our work. As a simple estimate, we can assess the quota of spin angular momentum necessary to spin the Be star progenitor up to its critical rate, and evaluate the impact of such spin-up on the final period of our models. For our models that reach a mass ratio at detachment as high as $m_{\mathrm{Be}}/m_{\mathrm{stripped}}\sim 13$ with a period of $P\sim 90$ days, we find that such final period would be reduced by only $\sim$ 7\%, if the Be star progenitor would spin up to critical at detachment.

Other angular momentum loss mechanisms than the isotropic re-emission mode could also have an influence on our results. Fully non conservative MT via isotropic re-emission requires much tighter initial orbital configurations than the conservative case (see Fig. \ref{fig:theory}), and this makes our models experience L2 overflow or inverse MT before forming a stripped + Be star system. However, larger initial periods are allowed if material is removed from the system carrying a specific angular momentum lower than that of the accretor. We decided to explore the case in which material is expelled from the system with a fixed fraction of the specific total orbital momentum, and we recomputed our grids as described in Appendix \ref{app:variation}. Our results show that, for the same amount of mass lost from the system, stable MT with this mode of angular momentum loss can achieve a target mass ratio at $P_{\mathrm{today}}$ more extreme than with the isotropic re-emission mode. However, the most extreme mass ratio achievable at $P_{\mathrm{today}}$ with this mode is $q_{\mathrm{r}}\sim 11.5$, therefore not succeeding in explaining HR6819's properties (see Appendix \ref{app:variation}). 

The total mass determination of $m_{\mathrm{tot,today}}=4.28\:M_{\odot}$ from \cite{klementVLTIGRAVITYEnables2025} is dependent on the distance measurement. They found $d=296.0\pm 8.0\:\mathrm{pc}$ from dynamical parallax, which to $2\sigma$ uncertainty allows for a total mass between $3.56\leq m_{\mathrm{tot,today}}\:(M_{\odot})\leq 5.00$, the range explored in this work. Following the calibration of \cite{Maiz2022} with an OB-type prior for the Gaia DR3 parallax, the distance can be much higher,  $d_{\mathrm{GAIA}}=375^{85}_{-58}\:\mathrm{pc}$. Considering that the total mass scales with the distance to the third power and $(d_{\mathrm{GAIA}}/d)^3\simeq 2.03$, the total mass determination could potentially shift to $\simeq 8.70\:M_{\odot}$. This much higher total mass could help explain the high luminosity of the system, as we could have a $\sim 8\:M_{\odot}$ Be star in orbit with a $\sim 0.5\:M_{\odot}$ stripped star (consistent with the typical masses of sdOBs, e.g. \citealt{Arancibia2024}), provided that we accept a possible inconsistency on the individual masses determinations from \cite{klementVLTIGRAVITYEnables2025}. However, a higher present-day total mass implies higher mass progenitors, which would in turn undergo L2 overflow, given how tight the admitted initial orbital periods are. We confirmed this prediction by producing grids similar to that of Fig. \ref{fig:grid} but relaxing the total mass constraint to match a present-day value of $\sim 8\:M_{\odot}$, and we found no solutions.

Lastly, one could think of a possible formation scenario that includes a common envelope episode in place of stable MT. This can be motivated by the relatively short period of HR6819 with respect to the other known (bloated) sdOB + Be binaries, which have orbits of order months. However, this relatively moderate separation still appears too wide to be the result of CE ejection, according to binary population synthesis studies (e.g. \citealt{han_origin_2003}). Nevertheless, the uncertainties in common envelope theory would make the outcomes largely unpredictable and we did not attempt any modeling. We can stress that highly non conservative MT would imply a much larger initial total mass than the present-day value. For case A systems, which is what is needed to achieve the lowest possible stripped star mass, unstable MT can be triggered at moderate to high initial mass ratios, but the absence of a defined core-envelope structure would make it harder for the binary to survive (e.g., \citealt{Ivanova2013}). A Case B or later MT phase offers a greater chance of survival and envelope ejection due to the presence of a compact core, but only stars with masses as low as $\sim 2.5\:M_{\odot}$ can produce a $\sim 0.27\:M_{\odot}$ helium core at the Terminal Age Main Sequence. Such a star could initiate the common envelope in a binary with a relatively moderate mass ratio, but the constraint on the present-day total mass would be hard to meet if the stripped star progenitor successfully ejects its $\sim 2.2\:M_{\odot}$ envelope.

HR6819 is the best constrained system with a puffed-up stripped star + Be star and, compared to the other similar known systems, exhibits the most extreme mass ratio. Such mass ratio of $m_{\mathrm{Be}}/m_{\mathrm{stripped}}=15.7\pm 1.1$ is also very extreme when compared with those of the sdOB + Be star system with constrained dynamical solutions ($\phi$ Persei \citealt{mourard_spectral_2015}; $\kappa$ Dra \citealt{Klement2022}; and other 6 in \citealt{klementCHARAArrayInterferometric2023}). Between these, HR2142 is the only one showing a more extreme median value for its mass ratio, with $m_{\mathrm{Be}}/m_{\mathrm{stripped}}=17.12\pm 2.11$, $m_{\mathrm{Be}}=17.6\pm 5.7\:M_{\odot}$ and $m_{\mathrm{stripped}}=1.03\pm 0.22\:M_{\odot}$ (see \citealt{klementCHARAArrayInterferometric2023}, their Table 5). Within these large uncertainties, this system may still fall within the range of mass ratios achievable with stable MT, given its wider orbital period of $P=80.8733\pm 0.0044$ days. Nevertheless, systems like HR6819 and HR2142 might both be pushing the stable MT channel to its limit, substantiating our finding that something is missing in our understanding of how stripped stars + Be star systems can form. Interestingly, \citealt{Lechien2025} also point out that both HR6819 and HR2142 are difficult to explain in their study, which assumes that systems evolve through case B stable MT, and state that they might have been formed by case A MT given their short initial periods and high accretion efficiency constraints. With this respect, our study of HR6819 complements theirs and further supports the problematic nature of the system, since we have investigated both case A and B interactions and found no consistent solution.

\begin{acknowledgements}
      The authors would like to thank Douglas Gies for reviewing the manuscript and giving helpful comments. AP acknowledges support from the FWO under grant agreement No. 11M8325N (PhD Fellowship), and K209924N, K223124N, K1A4925N (Travel Grants). PM and TS acknowledge support from the European Research Council (ERC) under the European Union’s Horizon 2020 research and innovation programme (grant agreement No. 101165213/Star-Grasp and 101164755/METAL). DP acknowledges financial support from the FWO in the form of a junior postdoctoral fellowship No. 1256225N. JSGM acknowledges having received funding from the European Research Council (ERC) under the Horizon Europe programme (Synergy Grant agreement No. 101071505: 4D-STAR). While partially funded by the European Union, views and opinions expressed are however those of the authors only and do not necessarily reflect those of the European Union or the European Research Council. Neither the European Union nor the granting authority can be held responsible for them. The resources and services used in this work were provided by the VSC (Flemish Supercomputer Center), funded by the Research Foundation - Flanders (FWO) and the Flemish Government.
\end{acknowledgements}

\bibliographystyle{aa}
\bibliography{My_Library.bib}

\begin{appendix}
\section{Details of \text{MESA} simulations}\label{sec:appMESA}
Our simulations are computed using version 24.03.1 of \text{MESA}. Our setup is similar to that of \cite{marchant_new_2016}, to which we refer for more details. We set the initial composition to correspond to the Galactic one from \cite{Brott2011}, for both stars. We used custom opacity tables computed from the OPAL project (\citealt{Iglesias_1996}) with solar-scaled metal abundances from \cite{Grevesse_1996}. 

Convection was modelled using the Ledoux criterion (\citealt{Ledoux1947}) within the standard mixing-length theory (\citealt{Bohm-Vitense1958}), with a mixing-length parameter $\alpha_{\mathrm{MLT}}=1.5$. Semiconvection was modelled according to \cite{Langer1983} with an efficiency parameter $\alpha_{\mathrm{sc}}=1$. For hydrogen burning cores, we include step-overshooting extending the convective region by 0.10 pressure scale height at the convective boundary (but see main text where we discussed the impact of varying this number). For convective cores after the main-sequence we include exponential overshooting (\citealt{Herwig2000}) with decay length of $f =$ 0.01. Thermohaline mixing is modeled as in \cite{Kippenhahn1980} with an efficiency parameter of $\alpha_{\mathrm{th}}=1$.

Stellar winds follow \cite{Yoon_2005}, with mass-loss rates for hydrogen-rich stars (with a surface helium abundance Y < 0.4) computed as in \cite{Vink2001}, while for hydrogen-poor stars (Y > 0.7) we use those from \cite{Hamann1995} multiplied by a factor of one tenth. In the range 0.4 < Y < 0.7, the rate is interpolated between the two.

Nuclear reaction rates in \text{MESA} are from JINA REACLIB \citep{Cyburt2010}, NACRE \citep{Angulo1999} and additional tabulated weak reaction rates \citet{Fuller1985, Oda1994, Langanke2000}. We use the simple networks provided with \text{MESA} \texttt{basic.net} for H and He burning, \texttt{co\_burn.net} for C and O burning, and \texttt{approx21.net} for later phases, with the inclusion of iron and calcium. 

Roche lobe radii in binary systems are computed using the fit of \citet{Eggleton1983}. Mass transfer rates in Roche lobe overflowing binary systems are computed with the contact scheme from \cite{marchant_new_2016}. This is an implicit scheme in which the mass transfer rate is iteratively adjusted to ensure that the donor star remains within its Roche lobe. The method can also accommodate contact phases where both stars overfill their Roche lobes, provided that mass loss through the second Lagrangian point L2 does not occur.

\section{Parameter variations}\label{app:variation}
\subsection{Reduced overshooting}
\begin{figure}
   \centering
   \includegraphics[width=0.5\textwidth]{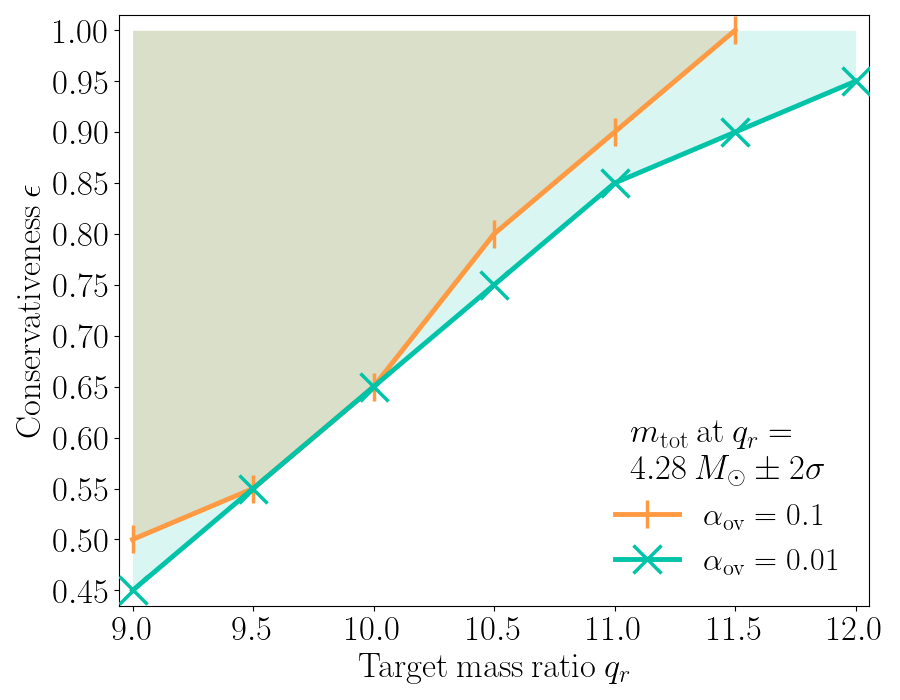}
      \caption{Comparison of our grid exploration for different conservativeness degree $\epsilon$ and different target mass ratios $q_{\mathrm{r}}$, for two different values of the overshooting parameter. The area is filled where a solution with $q_{\mathrm{detach}}\geq q_{\mathrm{r}}$ is found for a value of the present-day total mass $m_{\mathrm{tot,r}}$ within $2\sigma$ uncertainty. The scatter points along the respective lines indicate where the minimum value of $\epsilon$ that gives a solution is found.
              }
         \label{fig:Epsilon_overshooting_comparison}
   \end{figure}

All our simulations have been carried out with a step-overshooting scheme of the hydrogen burning convective core, extending its size by $0.1$ pressure scale heights. This was picked because lower overshooting values $\sim$ 0.1-0.2 have been proposed for lower-mass stars based on double-lined eclipsing binaries (\citealt{Claret2016}). To evaluate the impact of this choice, we re-computed our grids with a much lower value $\alpha_{\mathrm{ov}}=0.01$, to see if the limit on the maximum mass ratio at $P_{\mathrm{today}}$ achievable with stable MT would significantly shift. We show the result of the comparison in Fig. \ref{fig:Epsilon_overshooting_comparison}. Notice that we have also computed grids of simulations with $q_r=12.5$ and $0.50\leq \epsilon \leq 1.00$, spaced by $\Delta\epsilon =0.50$, but found no solution.

\subsection{Average orbital angular momentum}
\begin{figure}
   \centering
   \includegraphics[width=0.5\textwidth]{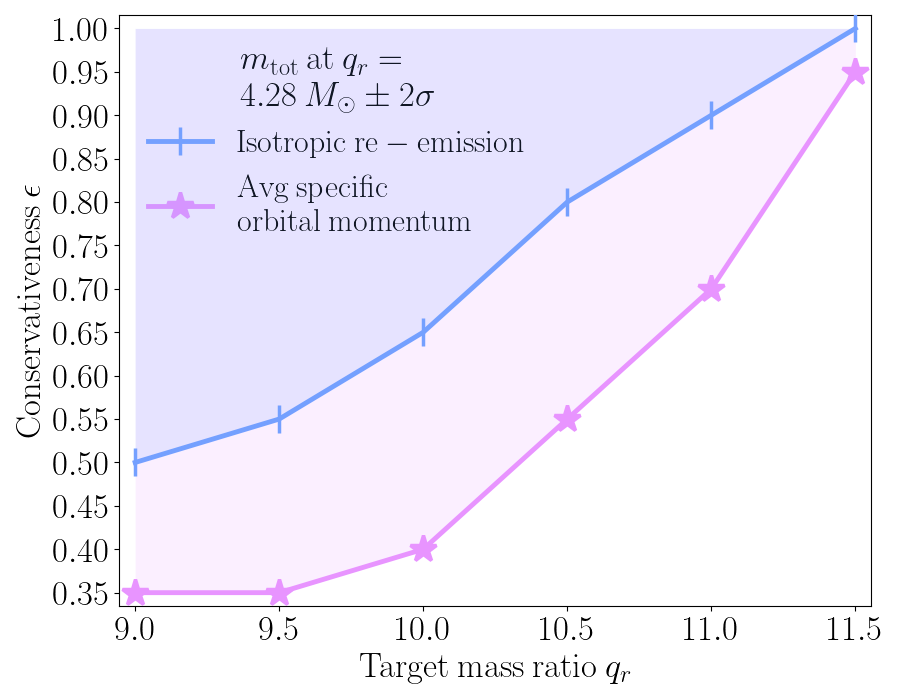}
      \caption{Similar to Fig. \ref{fig:Epsilon_overshooting_comparison}, but comparing two different angular momentum loss mechanisms: isotropic re-emission with efficiency $\beta=1-\epsilon$, and loss of a fixed fraction $\alpha_{\mathrm{avg}}=1-\epsilon$ of the specific total orbital momentum.}
         \label{fig:Epsilon_AM_comparison}
   \end{figure}

We introduce a fixed fraction $\alpha_{\mathrm{avg}}$ of material leaving the system with the average specific orbital momentum of the system (the total angular momentum of the orbit, divided by its total mass). Under these assumptions, one can derive another map between the initial orbital period $P_{\mathrm{i}}$ and the reference period $P_{\mathrm{r}}$:
\begin{equation}\label{eq:shrinkage_avg}
  \dfrac{P_{\mathrm{i}}}{P_{\mathrm{r}}}=\left(\dfrac{q_{\mathrm{r}}+1}{q_{\mathrm{i}}+1}\right)^{-\frac{3\alpha_{\mathrm{avg}}}{1-\epsilon}+1} \left(\dfrac{\epsilon q_{\mathrm{r}}+1}{\epsilon q_{\mathrm{i}}+1}\right)^{\frac{3\alpha_{\mathrm{avg}}}{1-\epsilon}-5}\left(\dfrac{q_{\mathrm{r}}}{q_{\mathrm{i}}}\right)^3\:,
\end{equation}
where the efficiency $\epsilon$ is defined as
\begin{equation}\label{eq:efficiency_avg}
    \epsilon=1-\alpha_{\mathrm{avg}}\:.
\end{equation}
Notice that the exponents show a factor $\alpha_{\mathrm{avg}}/(1-\epsilon)$: when MT is fully conservative (non conservative), this factor is zero (unity). We use this map combined with the total mass constraint in Eq. \ref{eq:total_mass} to determine initial conditions for a new set of grids, analogously to Sec. \ref{sec:methodology}. The results are shown in Fig. \ref{fig:Epsilon_AM_comparison}, compared with the isotropic re-emission mode. Grids of simulations with $q_r=12$ and $0.50\leq \epsilon \leq 1.00$, spaced by $\Delta\epsilon =0.50$, gave no solution.

\end{appendix}

\end{document}